\documentstyle[aps,prb,twocolumn,psfig]{revtex}

\draft
\begin{document}
\twocolumn[\hsize\textwidth\columnwidth\hsize\csname@twocolumnfalse%
\endcsname

\title{Magnetization in Molecular Iron Rings}

\author{B. Normand,$^{1,2}$ X. Wang,$^{3,4,5}$ X. Zotos,$^3$ and Daniel 
Loss$^2$}

\address{$^1$Theoretische Physik III, Elektronische Korrelationen und 
Magnetismus, Institut f\"ur Physik, Universit\"at Augsburg, 
D-86135 Augsburg, Germany}

\address{$^2$Departement f\"ur Physik und Astronomie, Universit\"at Basel, 
CH-4056 Basel, Switzerland}

\address{$^3$Institut Romand de Recherche Num\'erique en Physique des 
Materiaux (IRRMA), PPH-Ecublens, CH-1015 Lausanne, Switzerland}

\address{$^4$Max-Planck-Institut f\"ur Physik Komplexer Systeme, 
N\"othnitzerstra$\beta$e 38, D-01187 Dresden, Germany}

\address{$^5$Institute of Theoretical Physics, Chinese Academy of Sciences, 
P.O. Box 2735, Beijing 100080, People's Republic of China}

\date{\today}

\maketitle

\begin{abstract}

The organometallic ring molecules Fe$_6$ and Fe$_{10}$ are leading 
examples of a class of nanoscopic molecular magnets, which have been 
of intense recent interest both for their intrinsic magnetic properties 
and as candidates for the observation of macroscopic quantum coherent 
phenomena. Torque magnetometry experiments have been performed to 
measure the magnetization in single crystals of both systems. We 
provide a detailed interpretation of these results, with 
a view to full characterization of the material parameters. We present 
both the most accurate numerical simulations performed to date for ring 
molecules, using Exact Diagonalization and Density Matrix Renormalization 
Group techniques, and a semiclassical description for purposes of 
comparison. The results permit quantitative analysis of the variation of 
critical fields with angle, of the nature and height of magnetization and 
torque steps, and of the width and rounding of the plateau regions in both 
quantities.

\end{abstract}


\vspace{0.5cm}
]

\section{Introduction}

        The molecular iron ring, or ``ferric wheel'' systems Fe$_6$ and 
Fe$_{10}$\cite{rgcps} present an interesting subgroup of magnetic molecular 
clusters. Both materials have dominant antiferromagnetic (AF) coupling 
between spins $S = 5/2$ on each iron site, and have a ground state of 
total spin $S = 0$. The magnetic properties of such nanoscopic molecules 
result from the interplay of superexchange interactions between the atomic 
spins, dipolar coupling of the local moments, and on-site spin anisotropies 
arising from ligand configurations. In view of these complexities, and of 
the possibilities which exist\cite{rcajag} to influence the relative 
strengths of each interaction type through the non-Fe constituents of 
the molecules, a detailed understanding of the magnetic response is 
crucial. In appropriate parameter regimes, the ring systems are 
considered\cite{rldg,rcl} to be candidates for the observation of 
``macroscopic'' quantum phenomena, in the form of quantum coherent 
tunneling of the N\'eel vector.

        The magnetization $M(B)$ of single crystals of both Fe$_6$ and 
Fe$_{10}$ has recently been studied experimentally by cantilever torque 
magnetometry,\cite{rcja} using magnetic fields up to 23T, and at 
temperatures down to 0.45K. In the Fe$_6$ system, a selection of 
materials is available:\cite{rcajag} for Na:Fe$_6$ only the first 
magnetization step, to the $S = 1$ state and occurring at the field 
$B_{c1} \simeq 17$T, is accessible; for Li:Fe$_6$, $B_{c2} \simeq 21$T 
is also observed. For Fe$_{10}$, the fields $B_{c1}$ to $B_{c4}$, 
corresponding to the lowest four magnetization steps, may be measured. 
In a very recent analysis\cite{raeau} of Na:Fe$_6$, temperatures as low 
as 30mK were probed, and a temperature-independence of the step width 
observed, a result to which we will return later. In each material, 
$B_{c1}$ is found to be largest in the orientation where the magnetic 
field is applied along the axes of the rings (which we will denote as 
$\psi = 0$), and smallest when the field is directed in the plane of 
the rings ($\psi = \pi/2$). The variation between these extrema is to 
a good approximation a sinusoidal function of $2\psi$.

        Here we present a detailed analysis of the magnetization in 
these ferric wheels, considering the spacing and widths of the 
magnetization steps, the flatness of the plateaus, and the angular 
variation of the critical fields. The primary results are given by 
numerical simulations: we perform both the first Lanczos diagonalization 
of the full Fe$_6$ system, and the first Density Matrix Renormalization
Group (DMRG) studies of finite, $S = 5/2$ systems. Both techniques are 
exact at zero temperature for 6-site rings, and DMRG gives highly accurate 
results also for the 10-site case. As an aid to interpretation, we appeal 
in Sec. II to a semiclassical formulation for the molecular spin as a 
rigid-rotor model, and for the low-energy dynamics of the N\'eel 
vector.\cite{rcl} In Sec. III we compare this model with existing 
magnetization data to extract the relevant materials parameters. In 
Sec. IV we present the numerical simulations, and discuss the qualitative 
and quantitative significance of the results. Sec. V contains a summary 
and conclusions.

\section{Analytical description}

        We begin by considering a semiclassical description for the 
behavior of magnetic molecular rings, in order to deduce approximate 
materials parameters for the simulations. Justification for the 
applicability of the semiclassical model may be found in 
Ref.~\onlinecite{rcl}. The minimal starting Hamiltonian for the ring 
system is 
\begin{equation} 
H = J \sum_{i = 1}^N {\bf S}_i {\bf \cdot S}_{i+1} + \sum_{i = 1}^N U_i 
({\bf S}_i) + \hbar {\bf h \cdot} \sum_{i = 1}^N {\bf S}_i ,
\label{esh}
\end{equation}
where $N$ = 6 or 10, ${\bf S}_1 = {\bf S}_{N+1}$, and ${\bf h} = g \mu_B 
{\bf B} / \hbar$. $J$ is the AF superexchange interaction, $U$ contains 
intra-ring dipolar interactions and single-ion anisotropy terms, and the 
final term is the Zeeman coupling. For a ring geometry, it is easy to show 
that dipolar interactions within a single molecule contribute only an 
effective easy-axis spin anisotropy, which favors the axis normal to the 
ring plane. In addition, the environment of each Fe ion due to the 
surrounding ligand groups in general has a non-cubic point-group symmetry; 
for the majority of cases the on-site symmetry may be taken as uniaxial, 
and, again with the assistance of the ring geometry which allows\cite{rcl} 
radial anisotropy to be subsumed within the uniaxial terms, we will assume 
this form of anisotropy. In all systems known to date, this has been found 
to be weak, and of easy-axis orientation in the same sense as above. Thus 
to a reasonable approximation, the separate contributions to the $U$ term 
may be combined as a single, effective on-site anisotropy term $\sum_{i = 
1}^N - k_z S_{i,z}^2$, in which the sign of $k_z$ is chosen such that a 
positive value implies an easy axis for the orientation of the local spins. 
In Ref.~\onlinecite{rcajag} it was shown that the dipolar and single-site 
components of the effective $k_z$ may be attributed separately.

        The Hamiltonian (\ref{esh}) is mapped\cite{rcl} to a form of 
nonlinear $\sigma$ model for the staggered magnetization ${\bf n}$, 
by expressing the spins ${\bf S}_i$ in coherent state representation as 
$S {\bf \Omega}_i = (-1)^i S {\bf n}_i + {\bf l}_i$, and integrating over 
the uniform component ${\bf l}$. One obtains 
\begin{eqnarray} 
{\cal L}_E & = & N \left\{ \frac{\hbar^2}{8J} \left[ \dot{\bf n}^2 + i 
{\bf h \cdot} ({\bf n} \wedge \dot{\bf n}) - ( h^2 - ({\bf h \cdot n})^2)
\right] \right. \label{ele} \nonumber \\ & & \;\;\;\;\;\;\;\;\;\;\;\;\; 
\left. - k_z S^2 n_{z}^2 \right\} . 
\end{eqnarray}
Gradient terms $(\partial_x {\bf n})^2$, where $x$ is the spatial coordinate 
around the ring, may be neglected for small $N$, because for a small, 
closed system such as the ring geometry they represent significantly 
higher-lying energy states. This semiclassical description of the 
low-energy physics is also suitable for analyzing quantum coherence 
effects.\cite{rcl}

        The Hamiltonian formulation canonically conjugate to ${\cal L}_E$ 
(\ref{ele}) is deduced by considering the continuous spatial variable 
${\bf x} = R {\bf n}$, its conjugate momentum ${\bf p} = \partial {\cal 
L}_E / \partial \dot{\bf x}$, and the angular momentum ${\bf L} = {\bf x} 
\wedge {\bf p}$. In real time $t = - i \tau$ this is 
\begin{equation} 
H = \frac{2 J}{N \hbar^2} {\bf L}^2 + {\bf h \cdot L} - N k_z S^2 n_{z}^2 ,
\label{errh}
\end{equation}
which may be split in obvious notation into $H_0 ({\bf L}) + H_A$, the 
latter term denoting the anisotropy contribution. $H_0$ has eigenstates 
$|L,L_z \rangle = |l,m \rangle$, with $l = 0,1,2,\dots$ and $-l \le m \le 
l$, and eigenvalues $E_{l,m} = (2J / N) l(l+1) + g \mu_B B m$. Thus $H_0$ 
describes a staircase-like magnetization with steps occurring at the field 
values $B_{cn} = 4 J n / g \mu_B N$, where $n = 0,1,2,\dots$ is an integer. 
This behavior accounts directly for the gross features of the observed 
magnetization curves.\cite{rcja} Inclusion of the anisotropy term $H_A$ 
allows one to proceed to the effective spin Hamiltonian\cite{rbg} employed 
in Ref.~\onlinecite{rcja}, and to a discussion of the variation of the 
critical field with angle, $B_{cn} (\psi)$. In practice, this is used 
to deduce the anisotropy parameter $k_z$. In the presence of a single-ion 
anisotropy, the energy levels at zero field of the first (triplet) 
excitations are not degenerate, and $B_{c1}$ is given by the field where 
the $m = -1$ level crosses the singlet. For arbitrary $\psi$, the 
anisotropy term takes the form 
\begin{eqnarray}
H_A & = & - N k_z S^2 \cos^2 \psi \, n_{z}^2 - N k_z S^2 \sin^2 \psi \, 
n_{x}^2 \label{eha} \nonumber \\ & = & - N k_z S^2 \left( \cos^2 \theta 
\, \cos^2 \psi + \sin^2 \theta \, \sin^2 \phi \, \sin^2 \psi \right) , 
\end{eqnarray}
which describes simply the fact that components $n_z$ and $n_x$ profit from 
the easy axis as the ring is tilted around the field. $\theta$ and $\phi$ 
are the axial and azimuthal polar coordinates. The eigenvalues $E_A$ are 
evaluated using the spherical harmonic wavefunctions $Y_{l,m} (\theta,\phi)$. 
The results for the singlet and triplet levels are 
\begin{eqnarray}
E_{0,0} & = & - \frac{1}{3} N k_z S^2 (\cos^2 \psi + \sin^2 \psi) \; = \; 
- \frac{1}{3} N k_z S^2, \label{ehev} \nonumber \\ E_{1,0} & = & \frac{4 J}{N} 
- \frac{3}{5} N k_z S^2 \cos^2 \psi - \frac{1}{5} N k_z S^2 \sin^2 \psi,
\\ E_{1,\pm 1} & = & \frac{4 J}{N} - \frac{1}{5} N k_z S^2 \cos^2 \psi 
- \frac{2}{5} N k_z S^2 \sin^2 \psi \nonumber \\ & & \;\;\;\;\;\; 
\pm g \mu_B B . \nonumber
\end{eqnarray}
Comparison of $E_{0,0}$ with $E_{1,-1}$ gives the result 
\begin{equation} 
B_{c1} = \frac{4 J}{g \mu_B N} \left[ 1 + \frac{k_z N^2 S^2}{30 J} 
\left( 1 - \frac{3}{2} \sin^2 \psi \right) \right] , 
\label{ebc1}
\end{equation}
and thus a clear demonstration that the angle-dependence observed in both 
ferric wheel systems corresponds to an easy-axis anisotropy for the 
spin direction in the original model. For future reference, we define
as $B_{c1}^0 = 4 J / g \mu_B N$ the field value where the $S = 0 
\rightarrow 1$ transition would occur in the absence of magnetic 
anisotropy, whence Eq.~(\ref{ebc1}) may be reexpressed as
\begin{equation} 
B_{c1} = B_{c1}^0 \left[ 1 + \frac{1}{15} \lambda_{c1} \left( 1 - 
\frac{3}{2} \sin^2 \psi \right) \right] , 
\label{ebc2}
\end{equation}
where $\lambda_{c1}$ is the value of the anisotropy-field ratio\cite{rcl} 
$\lambda = 8 k_z J S^2 / (g \mu_B B)^2$ at $B = B_{c1}^0$. Finally, by 
following the same procedure as in Eq.~(\ref{ehev}) for the wave function 
$Y_{2,-2}(\theta,\phi)$, one obtains 
\begin{eqnarray} 
E_{2,-2} & = & \frac{12 J}{N} - \frac{1}{7} N k_z S^2 \cos^2 \psi 
- \frac{3}{7} N k_z S^2 \sin^2 \psi \label{ehev2} \nonumber \\
& & \;\;\;\;\;\;\;\; - 2 g \mu_B B , 
\end{eqnarray}
and thus 
\begin{equation} 
B_{c2} = 2 B_{c1}^0 \left[ 1 + \frac{1}{70} \lambda_{c1} \left( 1 - 
\frac{3}{2} \sin^2 \psi \right) \right] , 
\label{ebc3}
\end{equation}
It is evident that in the presence of anisotropy, the widths of the 
magnetization steps are no longer equal. 

Before leaving the rigid-rotor model (Eq.~(\ref{errh})), we may employ 
it to make a valuable qualitative observation concerning the nature of 
the magnetization. The situation of most interest for quantum transitions 
between low-energy states of similar energy is when the field is applied 
perpendicular to the axis of the anisotropy. In the real materials, this 
anisotropy is rather weak, and it would be most appropriate to seek 
quantum coherent phenomena in the high-field regime of 
Ref.~\onlinecite{rcl}. If the field is dominant and sets the quantization 
axis ($\hat{z}$) of the rotor, the on-site anisotropy term takes the form 
$ - \sum_i k_z S_{x}^2 = - {\textstyle {\frac{1}{4}}} \sum_i k_z (S_{+} + 
S_{-})^2$, where $S_{\alpha}$ is a spin operator acting within the $S = 
5/2$ manifold on each site. The anisotropy, considered as a perturbation, 
may change the total spin state of the system only by $\Delta S = 0, \pm 
2$. All level crossings between eigenstates of the rigid rotor will then 
be true crossings when $\Delta m = \pm 1, \pm 3, \dots$, but will become 
anticrossings due to the mixing perturbation when $\Delta m = \pm 2, \pm 
4, \dots$ Because the ground-state level alterations always involve 
$\Delta m = 1$, we thus obtain a simple understanding for the fact that 
all of the magnetization steps will have vertical portions for the starting 
Hamiltonian (\ref{esh}), whereas the energy level splitting $E_{01} = E_1 - 
E_0$ will be not a sawtooth-shaped function but a rounded one.\cite{rcl} 
In principle, the anisotropy term may be reexpressed in terms of linear 
combinations of operators $L_{\alpha}$ for the rigid rotor, allowing one 
to obtain analytical expressions for the numerical results to follow.

\section{Experimental comparison}

The results of the previous section may be applied directly to the 
experimental data of Refs.~\onlinecite{rcja} and \onlinecite{rcajag}. 
The variation of the critical fields for the lowest magnetization steps 
with the angle $\psi$ between the applied field and the ring axis has 
been measured to high accuracy by torque magnetometry, and is reproduced 
in Figs.~4 (Na:Fe$_6$ and Fe$_{10}$) and 3 (Li:Fe$_6$), respectively, of 
these works. The mean value of the field at which the step occurs gives 
the antiferromagnetic superexchange coupling $J$, which is responsible 
for maintaining the $S = 0$ ground-state spin configuration at zero field. 
The variation of the critical field values $B_c$ between their extrema at 
$\psi = 0$ and $\psi = \pi/2$ gives the effective anisotropy parameter 
$k_z$ for each material. In the case of Li:Fe$_6$, the two steps which 
have been characterized give an independent verification of the consistency 
of the rigid-rotor description over a wider field range. 

\medskip
\noindent
\begin{tabular}{c|c|c|c|c|c}
& $B_{c1}^0$ & $J$ & $\Delta B_{c}$ & $k_z/J$ & $\theta$ \\ \hline
Na:Fe$_6$ $\,$ & $\,$ 16.32T $\,$ & $\,$ 32.77K $\,$ & $\,$ 2.49T $\,$ 
& $\,$ 0.0136 $\,$ & $\,$ 0.025 \\ 
Fe$_{10}$ $\,$ & $\,$ 4.65T $\,$ & $\,$ 15.56K $\,$ & $\,$ 1.28T $\,$ 
& $\,$ 0.0088 $\,$ & $\,$ -0.305 \\ 
Li:Fe$_6$ $\,$ & $\,$ 10.38T $\,$ & $\,$ 20.83K $\,$ & $\,$ 0.620T $\,$ 
& $\,$ 0.0053 $\,$ & $\,$ 0.011 \\ 
Li:Fe$_6$ $\,$ & $\,$ 10.30T $\,$ & $\,$ 20.68K $\,$ & $\,$ 0.328T $\,$ 
& $\,$ 0.0066 $\,$ & $\,$ -0.091 
\end{tabular}
\medskip

Fits of the data to Eqs.~(\ref{ebc1}) or (\ref{ebc2}) for the first steps, 
and to Eq.~(\ref{ebc3}) for the second step in Li:Fe$_6$, yield the values 
of $J$ and $k_z$ shown in Table 1. The parameters in the last row are 
taken from $B_{c2}$ for Li:Fe$_6$, and show quite satisfactory consistency 
with the values deduced from the first step. The results are illustrated 
in Figs.~1 and 2. The angular offset in the final column of the table is 
an alignment parameter concerning the experimental orientations of crystal 
and goniometer, and is not important for the intrinsic physics. This angle 
is quite large in Fe$_{10}$ (Fig.~1(b)), where the crystal structure 
contains rings of two different 
orientations, which are observed\cite{rcja} in the same projection on 
the field direction. The fitting procedure employed here bypasses the 
definition of an intermediate, effective model with parameters $D_i$; 
the results of Table 1 agree well with those of the original 
works,\cite{rcja,rcajag} which were extracted by different methods, 
and with the only other known torque measurements.\cite{rhapc} They 
are also fully consistent with the energy scales deduced from analyses 
of low temperature specific heat for Na:Fe$_6$ and Fe$_{10}$,\cite{ralcc}
and additionally for Li:Fe$_6$.\cite{ralc}

\begin{figure}[hp]
\centerline{\psfig{figure=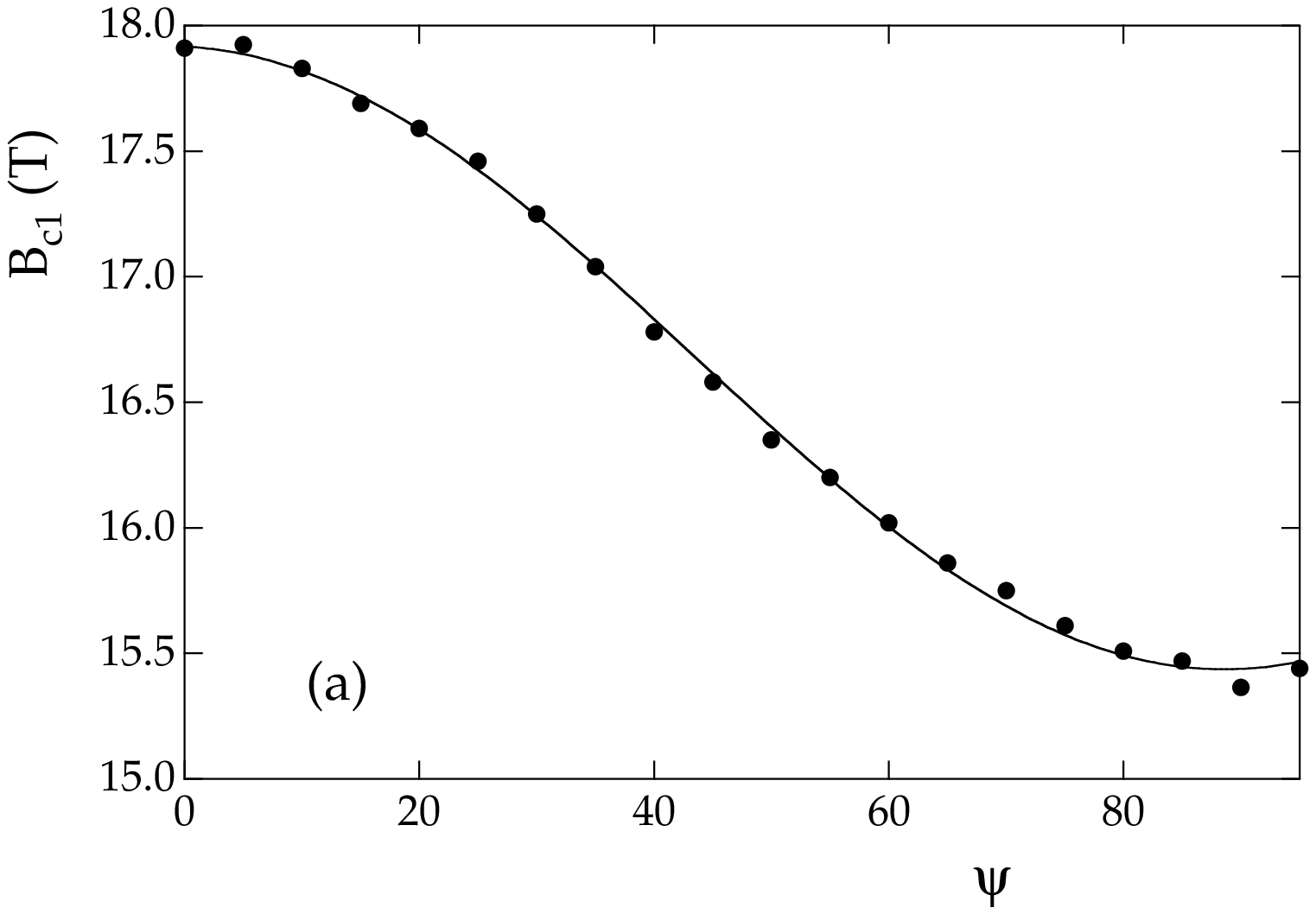,height=4.7cm,angle=0}}
\medskip
\centerline{\psfig{figure=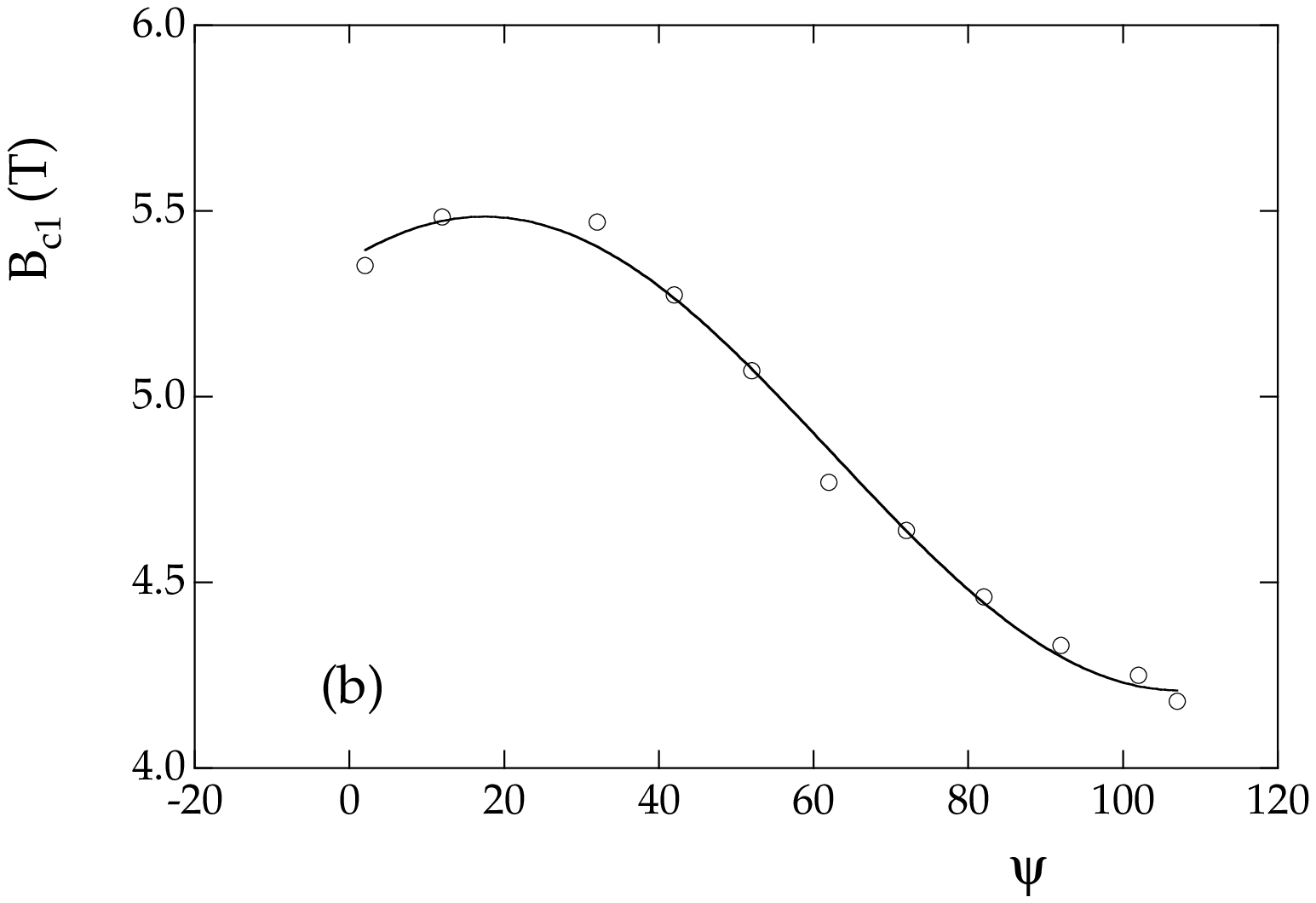,height=4.8cm,angle=0}}
\medskip
\caption{Angular variation of critical field $B_{c1}$ for first 
magnetization step in (a) Na:Fe$_6$ and (b) Fe$_{10}$. Data points 
are from Ref.~\protect\onlinecite{rcja}, and solid lines are fits 
using Eq.~(\ref{ebc1}) with the parameters in the Table. }
\end{figure}

\begin{figure}[hp]
\centerline{\psfig{figure=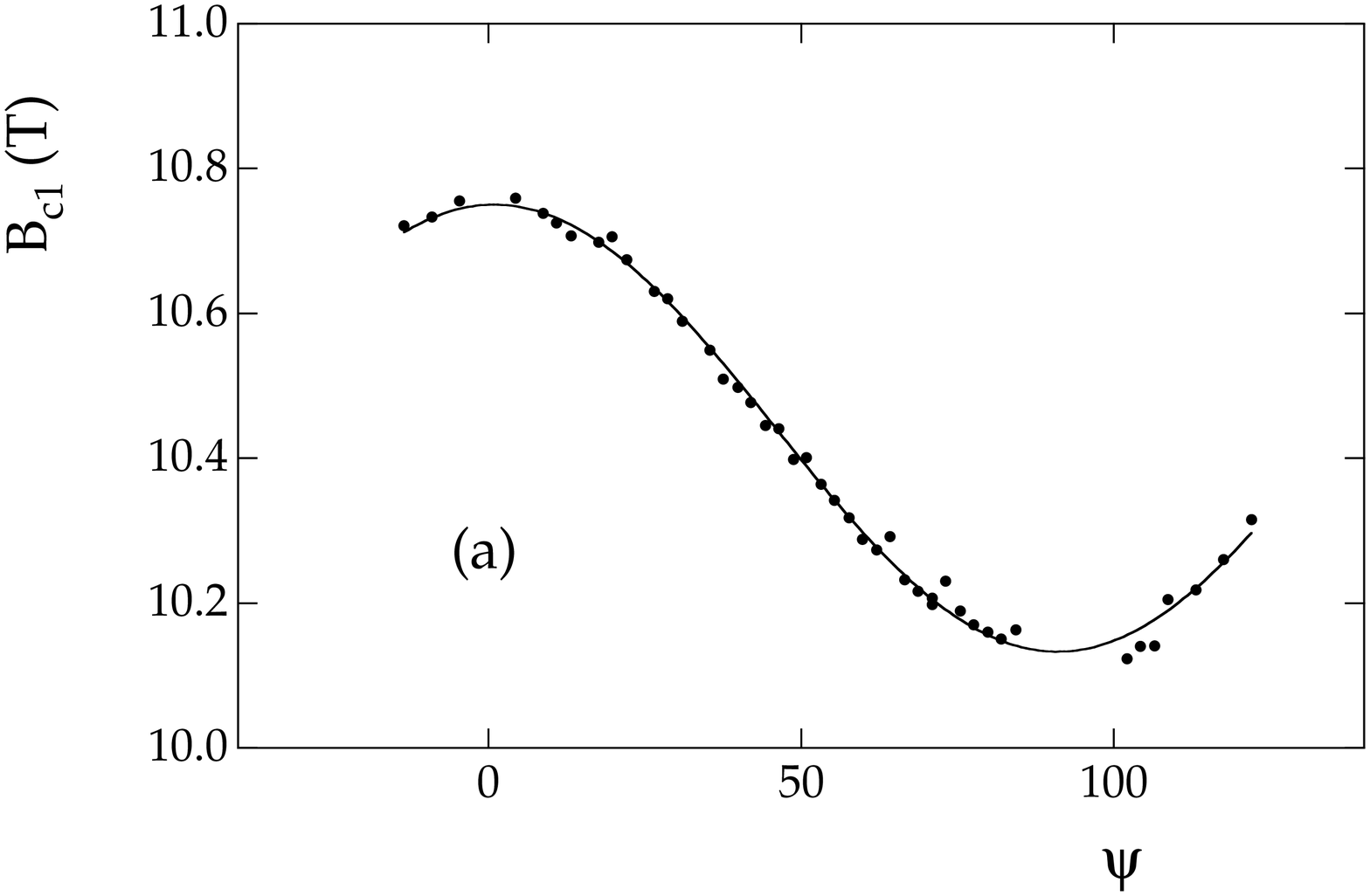,height=4.8cm,angle=0}}
\medskip
\centerline{\psfig{figure=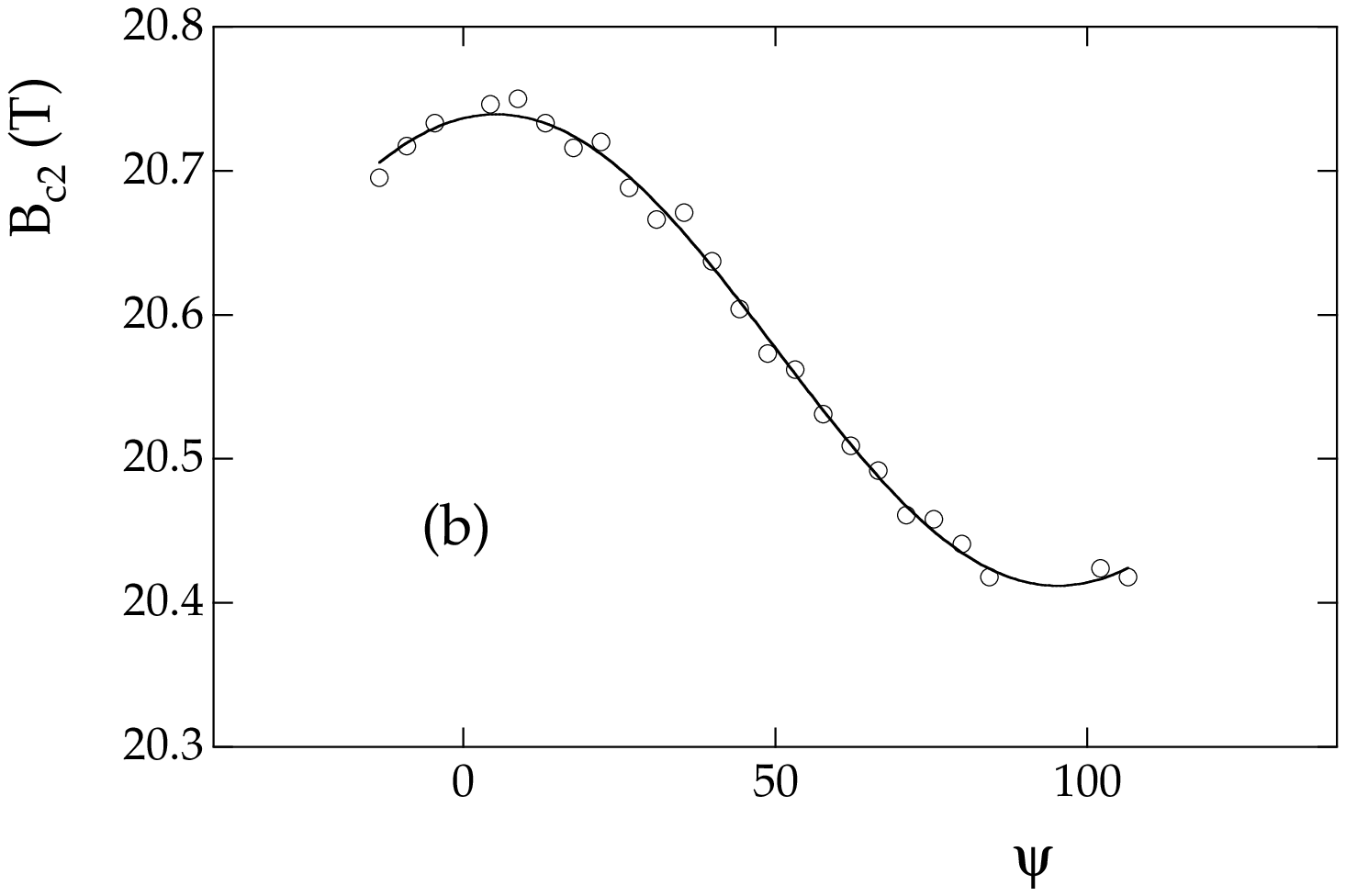,height=4.86cm,angle=0}}
\medskip
\caption{Angular variation of critical fields $B_{c1}$ (a) and $B_{c2}$
(b) in Li:Fe$_6$. Data points are from Ref.~\protect\onlinecite{rcajag}, 
and solid lines are fits using Eqs.~(\ref{ebc1}) and (\ref{ebc3}) with 
the parameters in the Table. }
\end{figure}

\section{Numerical Simulations}

With this semi-classical background for the leading-order effects, we 
turn now to fully quantum mechanical simulations of the model (\ref{esh}),
with the minimal, uniaxial form of the U term 
\begin{equation} 
H = J \sum_{i = 1}^N {\bf S}_i {\bf \cdot S}_{i+1} - \sum_{i = 1}^N k_z 
S_{i,z}^2 + \hbar {\bf h \cdot} \sum_{i = 1}^N {\bf S}_i .
\label{emmh}
\end{equation}
We work with arbitrary field angle and strength, and with focus on the 
experimental parameter regimes for Na:Fe$_6$ and Fe$_{10}$ (Table 1). 

\subsection{Exact Diagonalization}

For Fe$_6$, $N = 6$ spins $S = 5/2$ require dealing with $(2S + 1)^N = 
46656$ basis states. This Hamiltonian matrix is eminently accessible by 
the Lanczos exact diagonalization (ED) technique, which provides the 
lowest eigenvalues and their corresponding eigenstates. For any angle 
of the field to the ring plane, the magnetization is given by 
\begin{equation} 
M = \frac{\partial E}{\partial B} = \langle S_z \rangle \cos \psi + 
\langle S_x \rangle \sin \psi ,
\label{emfd}
\end{equation}
where $\langle S_{\alpha} \rangle = \sum_{i = 1}^N \langle S_{i,\alpha} 
\rangle$ is the ground-state expectation value of component $\alpha$ 
of the total-spin operator. The results are shown in Fig.~3(a) 
for Fe$_6$, with $k_z^{\rm Na} = 0.0136J$, and for angles $\psi = 0$, 
$\pi/4$, and $\pi/2$. The magnetization displays clearly the stepped 
form, with the step position varying as a function of field angle, which 
is observed in experiment and expected from the rigid-rotor model 
(\ref{errh}) of Sec. II. From the location of the steps in field at 
each angle, we find that the model (Eqs.~(\ref{ebc2},\ref{ebc3})) is 
in quantitative agreement with simulations at the 3-5\% level. Given 
its inherent approximations, the rigid-rotor description is remarkably 
satisfactory in the experimentally relevant regime of weak anisotropy 
which is considered here.

\begin{figure}[hp]
\centerline{\psfig{figure=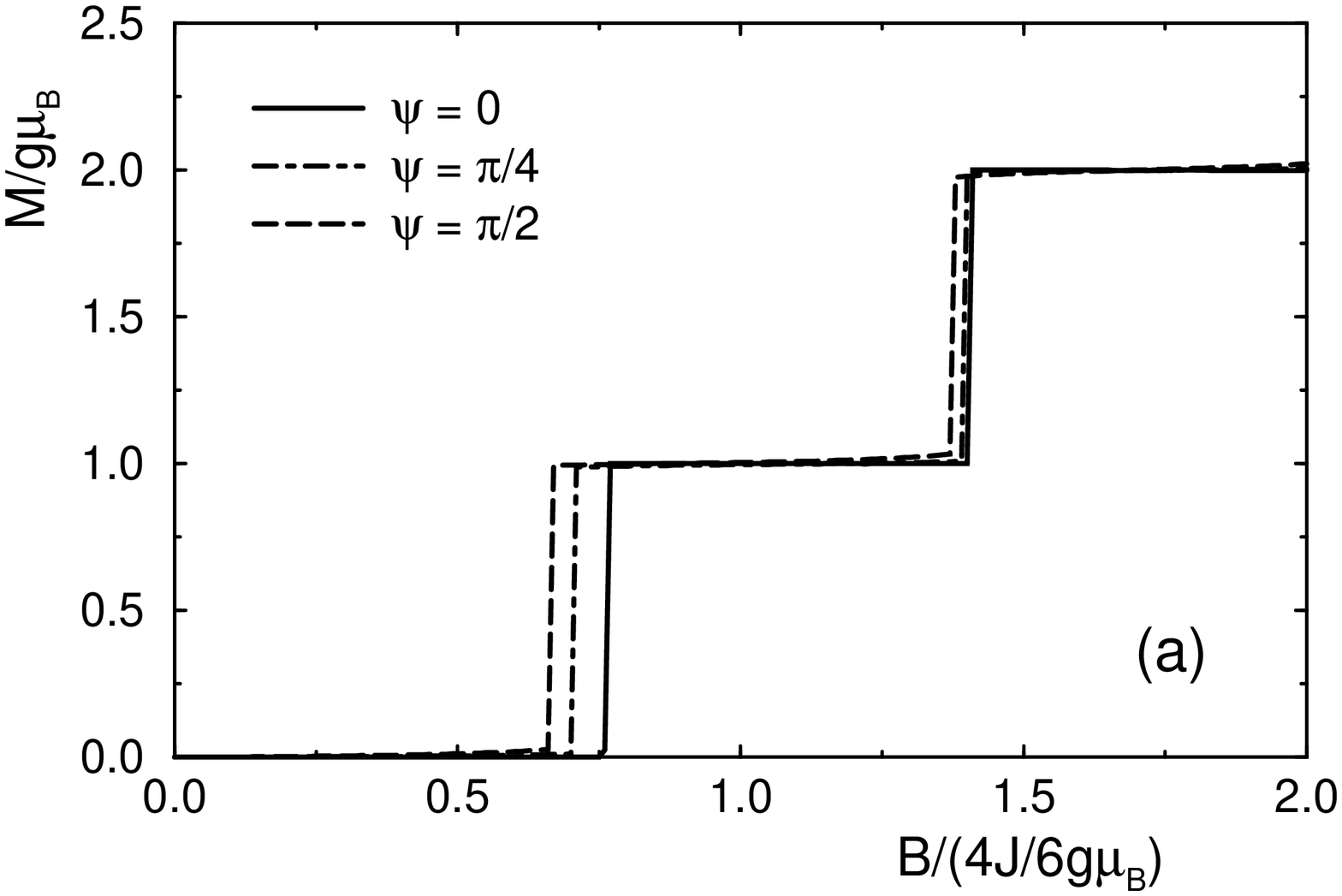,height=5cm,angle=270}}
\centerline{\psfig{figure=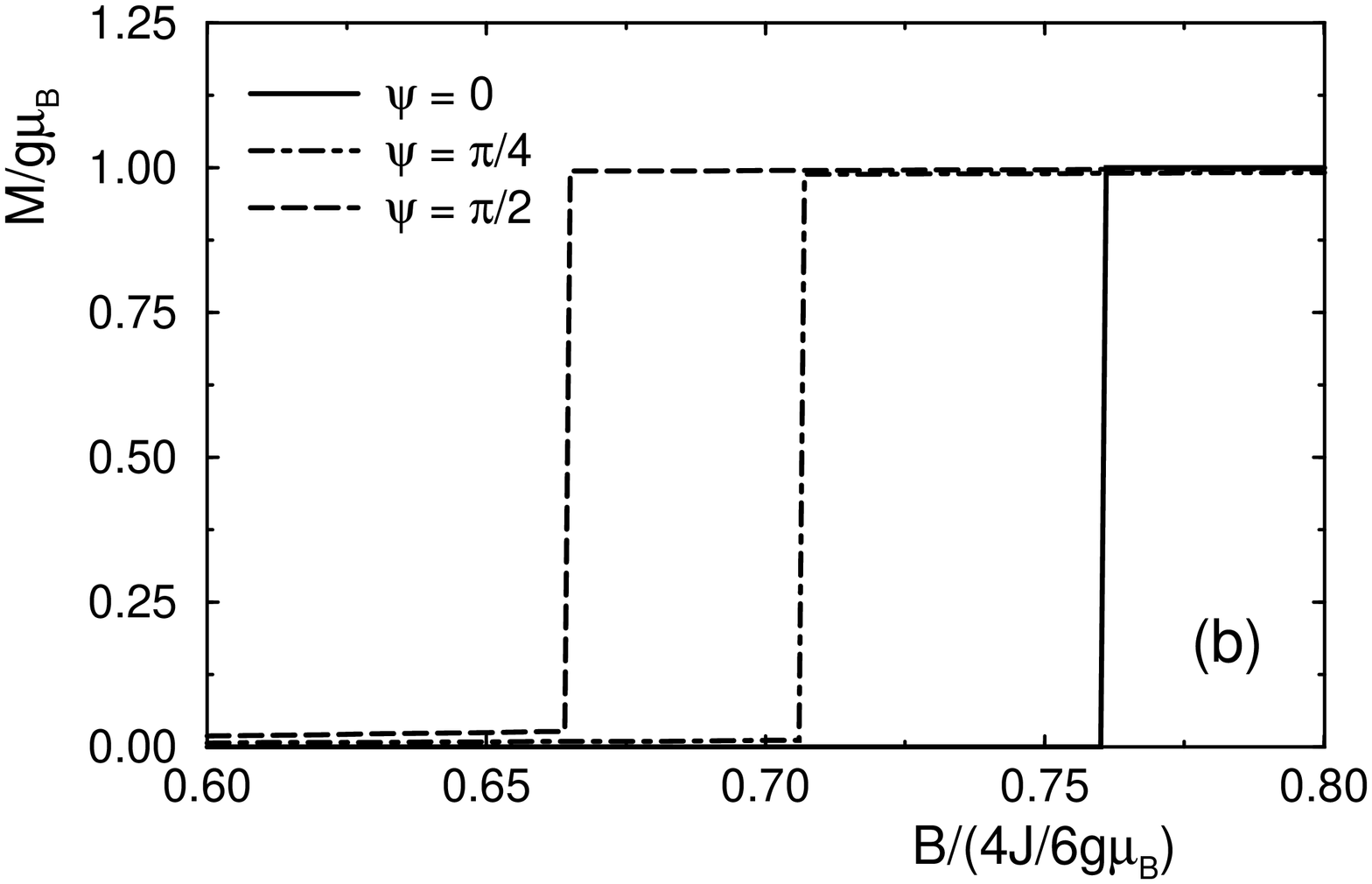,height=5cm,angle=270}}
\medskip
\caption{(a) Magnetization $M(B)$ for Na:Fe$_6$ from ED, for field angles 
$\psi = 0$, $\pi/4$, and $\pi/2$. (b) Magnification of $S = 0 \rightarrow 
1$ step region, illustrating anisotropy-induced variation in $B_{c1}$, 
and vertical nature of step. }
\end{figure}

The field resolution in these simulations was 0.001 at the steps in the 
units given. One observes that the magnetization steps are vertical, as 
shown in Fig.~3(b) around the first step. However, for angles $\psi \ne 
0$ they do not have the full height $\Delta M = 1$. This feature, and 
the curvature of the magnetization ``plateaus'', are illustrated more 
clearly in Fig.~4(a) for the same angles, but with the anisotropy 
taken as $k_z = 2 k_z^{\rm Na}$. They may be understood as a natural 
consequence of zero-point transitions between the nearly-degenerate 
energy levels of different spin sectors $S_z$ close to the step field 
$B_{c1}$. These are enhanced over a broader region of field around any 
step $B_{cn}$ as the field angle approaches $\psi = \pi/2$, and the 
system realizes more closely the optimal geometry\cite{rcl} of two 
degenerate spin states corresponding to the N\'eel vector ${\bf n}$ 
lying along $\pm {\hat z}$. Fig.~4(b) shows the magnetization curve 
for geometry $\psi = \pi/2$, for three anisotropy values $k_z = 0$, 
$k_{z}^{\rm Na}$, and $2k_{z}^{\rm Na}$. It is clear that in the 
absence of the on-site anisotropy term no quantum transitions may 
occur, because all energy levels are degenerate with no barrier. 
Deviations of the critical fields from their unperturbed values 
$B_{cn}^0$, and also the rounding of the plateaus, increase linearly 
with $k_z$. Plateau rounding effects are contained in the semiclassical 
theory \cite{rcl}, and may be compared at the qualitative level. 

\begin{figure}[hp]
\centerline{\psfig{figure=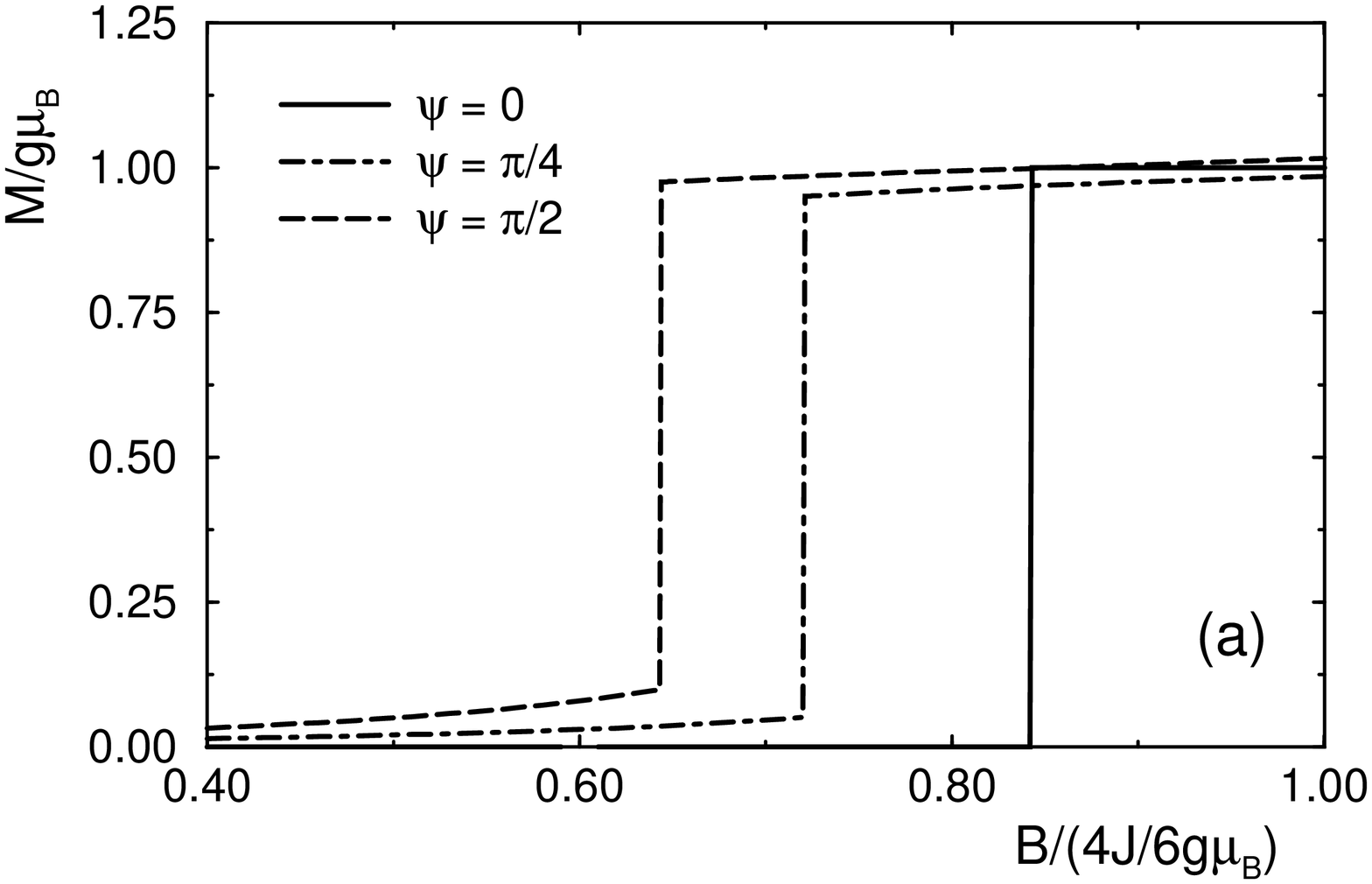,height=5cm,angle=270}}
\centerline{\psfig{figure=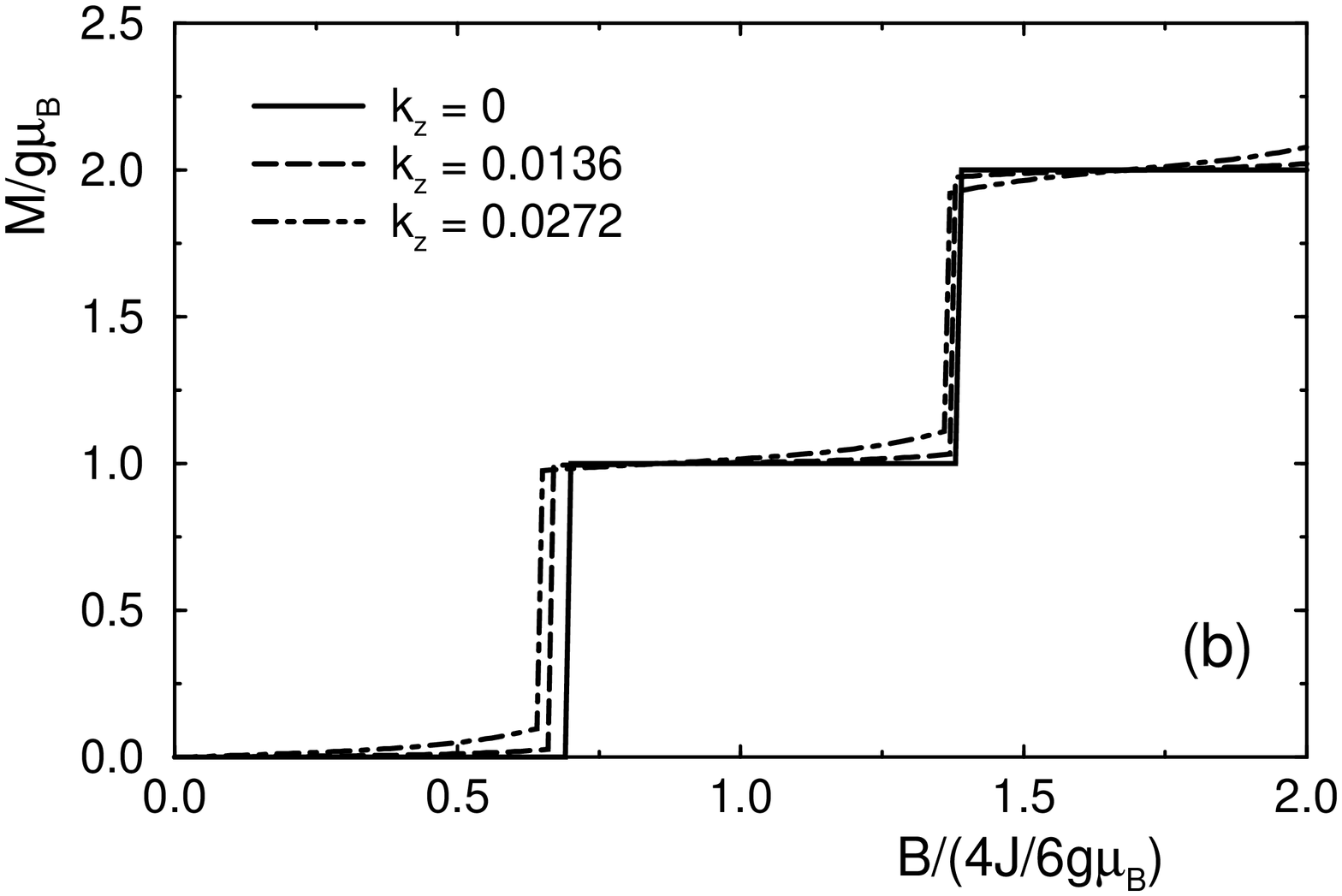,height=5cm,angle=270}}
\medskip
\caption{Magnetization $M(B)$ for an Fe$_6$ ring from ED, (a) for field 
angles $\psi = 0$, $\pi/4$, and $\pi/2$ at anisotropy $k_z$ = 0.0272 
(twice that in Na:Fe$_6$), and (b) with field angle $\psi = \pi/2$ and 
anisotropy values $k_z = 0$, $k_{z}^{\rm Na}$, and $2k_{z}^{\rm Na}$. }
\end{figure}

In Fig.~5(a) is shown the low-field torque ,
\begin{equation} 
T_y = - g \mu_B B ( \langle S_x \rangle \cos \psi - \langle S_z \rangle 
\sin \psi) . 
\label{et}
\end{equation}
For the geometry $\psi = \pi/4$ (illustrated), where 
the torque is maximal, it takes the form of a simple difference between 
the $x$ and $z$ spin components. As shown in Fig.~5(b), at higher fields 
this quantity is maximal around the 4th plateau, then decreases, changes 
sign, and adopts a negative constant value beyond saturation at the 
fifteenth plateau. Such behavior may be understood from the 
asymmetric field evolution of the spin components in the presence of a 
finite easy-axis anisotropy benefiting one of them. The torque curve, 
which is directly experimentally accessible, is thus characterized by 
vertical steps at the magnetization jumps, doubly rounded plateaus, and 
a distinctive variation in step heights. The first two features may be 
expected to be visible at the lowest temperatures available in Fe$_6$ 
systems, but the evolution of step heights would be hard to measure 
because of the high fields involved. 

\begin{figure}[hp]
\centerline{\psfig{figure=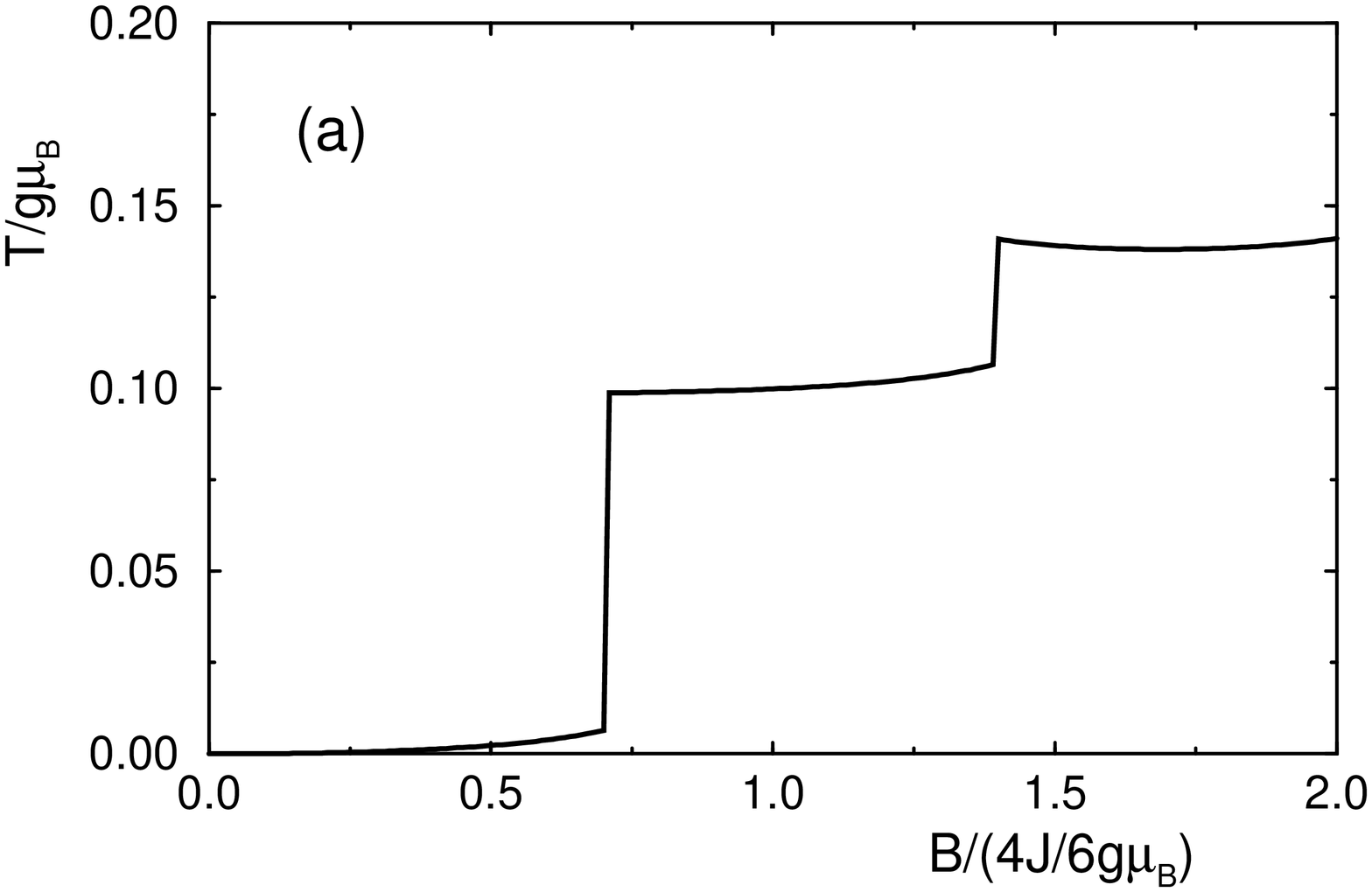,height=5cm,angle=270}}
\centerline{\psfig{figure=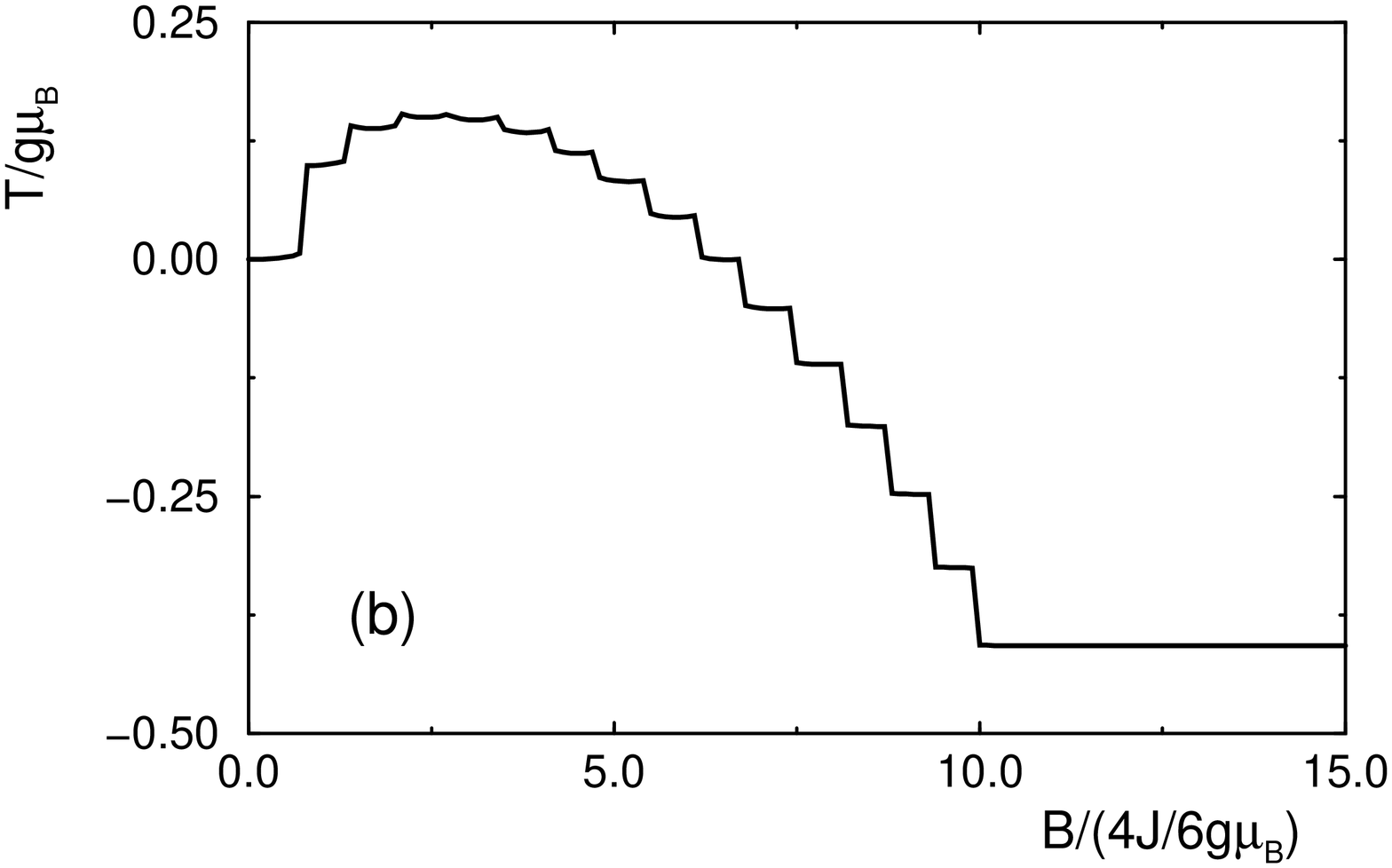,height=5cm,angle=270}}
\medskip
\caption{Torque curve of Na:Fe$_6$ for $\psi = \pi/4$. (a) Detail at 
low fields; (b) full field range. } 
\end{figure}

\subsection{Density Matrix Renormalization Group}

The DMRG technique\cite{rpwkh} is a powerful method for simulating the 
low-energy 
properties of one-dimensional systems, which can be much larger than 
those accessible by ED. Its extensions to periodic systems, and to high 
spins such as $S = 5/2$, are conceptually straightforward, but require 
considerable numerical effort due to the large spin degree of freedom 
and, for anisotropic ring systems in a field, the absence of symmetries. 
In our numerical calculations, we retain at least 80 states, and use 
a finite-size algorithm with 3-5 sweeps for each case. With these 
conditions, the results are altered by less than 1 part in 10$^3$ on 
adding more states or more sweeps. We present here the 
first results of simulations of this kind (Figs.~6-7). As for the ED 
results above, these are $T = 0$ data, and show essentially the exact 
magnetization and torque response of 10-site rings for the minimal 
model of Eq.~(\ref{esh}). Because the DMRG method applied to a 
sufficiently small system (up to 8 sites) involves a very similar 
truncation to Lanczos ED, the results of simulations on Na:Fe$_6$ are 
identical to Figs.~3-5; this provides a useful check of both techniques, 
and obviates further discussion of this system by DMRG. 

Fig.~6 shows the magnetization curve of Fe$_{10}$ for the physical 
parameter value $k_{z}^{10} = 0.0088$. The qualitative features of 
step location and plateau rounding with angle which are visible in the 
first three steps (Fig.~6(a)) are as in Figs.~3 and 4. Their relative 
sizes are similar to Na:Fe$_6$ because the smaller value of $k_z$ is 
offset by the larger value of $N$. Indeed, the plateau rounding at 
$\psi = \pi/2$ is found to be more pronounced for Fe$_{10}$ than for 
Na:Fe$_6$, while, as emphasized above, the step remains vertical; the 
field increment in Fig.~6(a) is 0.0125 in the units displayed. Fig.~6(b) 
shows the magnetization response at high fields, which has 25 steps 
corresponding to the $S_z$ manifolds available between zero and full 
saturation.

\begin{figure}[hp]
\centerline{\psfig{figure=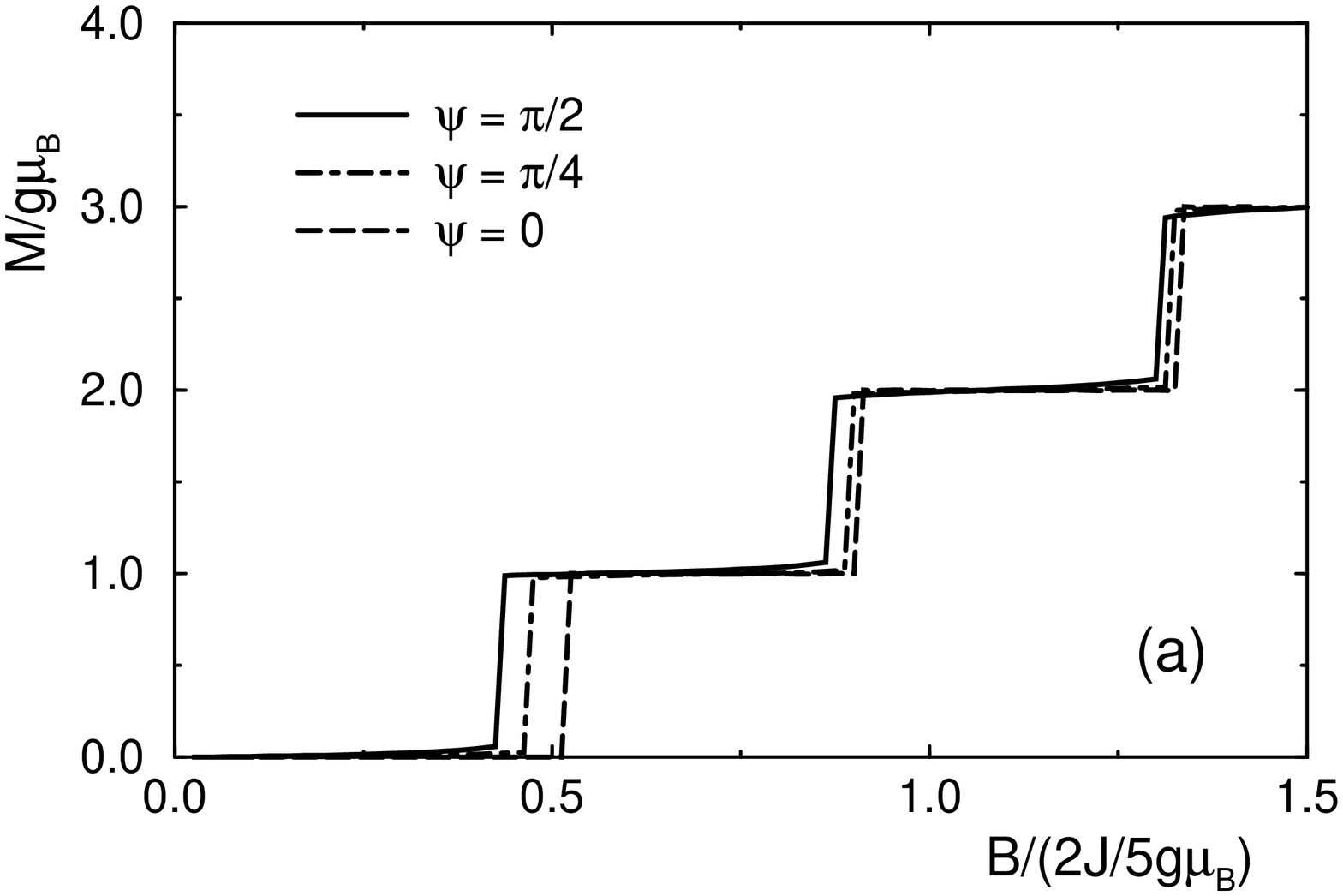,height=5cm,angle=270}}
\centerline{\psfig{figure=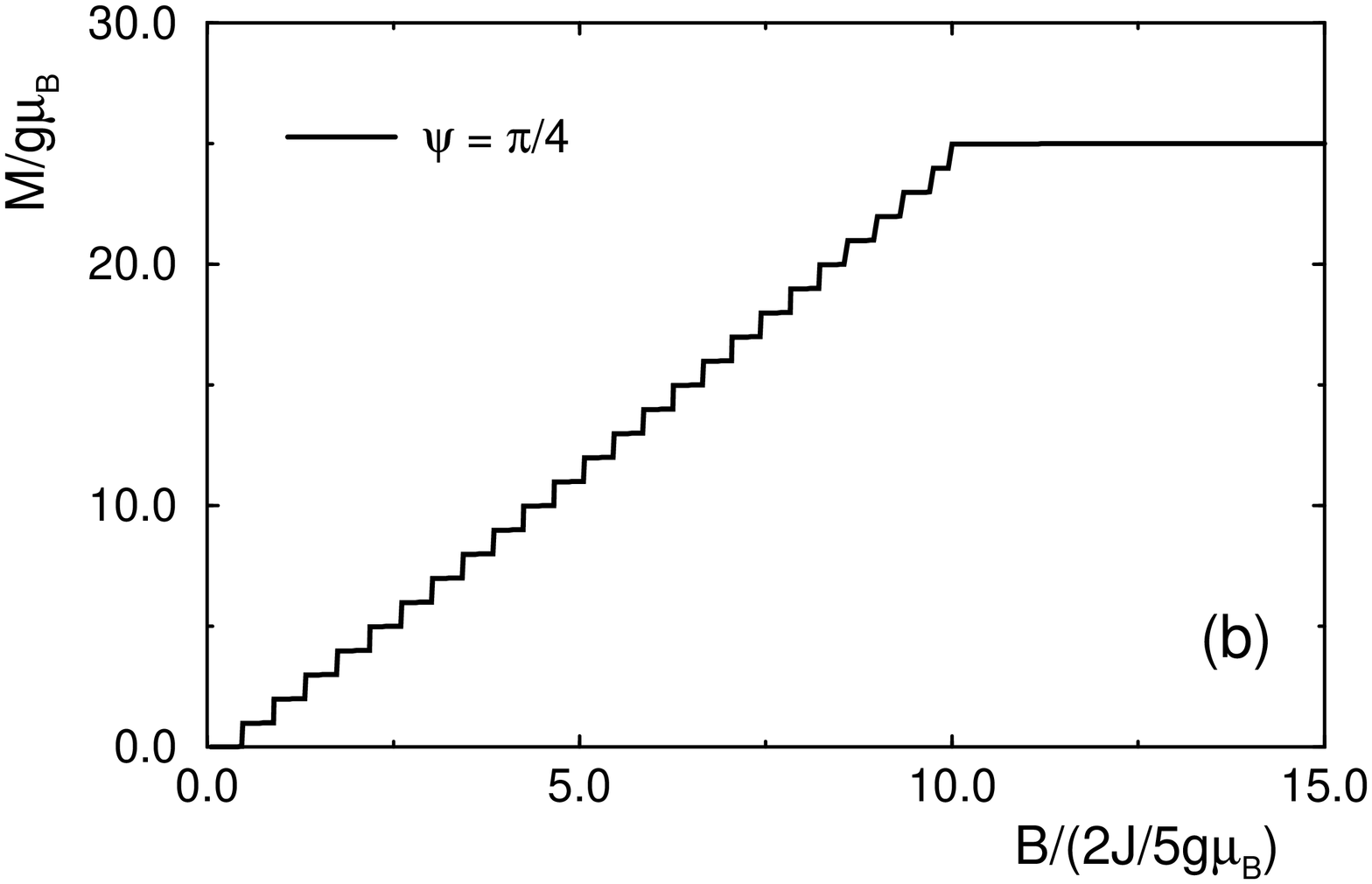,height=5cm,angle=270}}
\medskip
\caption{(a) Magnetization $M(B)$ for Fe$_{10}$ from DMRG, for field angles 
$\psi = 0$, $\pi/4$, and $\pi/2$. (b) $M(B)$ for field angle $\pi/4$ over 
the full field range, illustrating the 25 steps to saturation. } 
\end{figure}

Fig.~7 illustrates the torque for Fe$_{10}$ at $\psi = \pi/4$. The 
qualitative similarity between Fig.~7(a) the data of Ref.~\onlinecite{rcja} 
for the evolution of step heights is worthy of note, although the data at 
0.45K have too much thermal broadening for meaningful comparison with the 
predictions for critical field locations and plateau rounding. However, 
because the critical field in this system is four times smaller than in 
Na:Fe$_6$, the possibility arises of a detailed experimental 
characterization of this curve over the first five plateaus.\cite{raeau}
Following the torque response to high fields (Fig.~7(b)) shows that 
the maximal value appears around the sixth plateau, and saturation after 
the 25th. 

\section{Discussion}

The rigid-rotor model of Sec. II corresponds to the kinetic limit, 
of a particle moving in a weak potential, and thus is quite suitable for 
the real Fe$_6$ and Fe$_{10}$ systems where the anisotropy $k_z$ is weak, 
and thus the barrier height is small. In addition to the energy levels 
and step positions discussed in Sec. III, the Hamiltonian (\ref{errh}) may 
be extended in second-order perturbation theory to give a qualitative 
description of the rounding of torque and magnetization plateaus. By 
contrast, the Lagrangean of Eq.~(\ref{ele}) is most appropriate for the 
tight-binding limit of a particle in a strong potential, and for the 
semi-classical description of tunneling dynamics. This formulation gives 
a useful qualitative account of the energy level splitting, and of tunneling 
processes which result in plateau rounding, but is not applicable to the 
experimental curves at low temperatures. The numerical simulations remain 
the most reliable source of quantitative torque and magnetization data for 
the real system parameters. 

\begin{figure}[hp]
\centerline{\psfig{figure=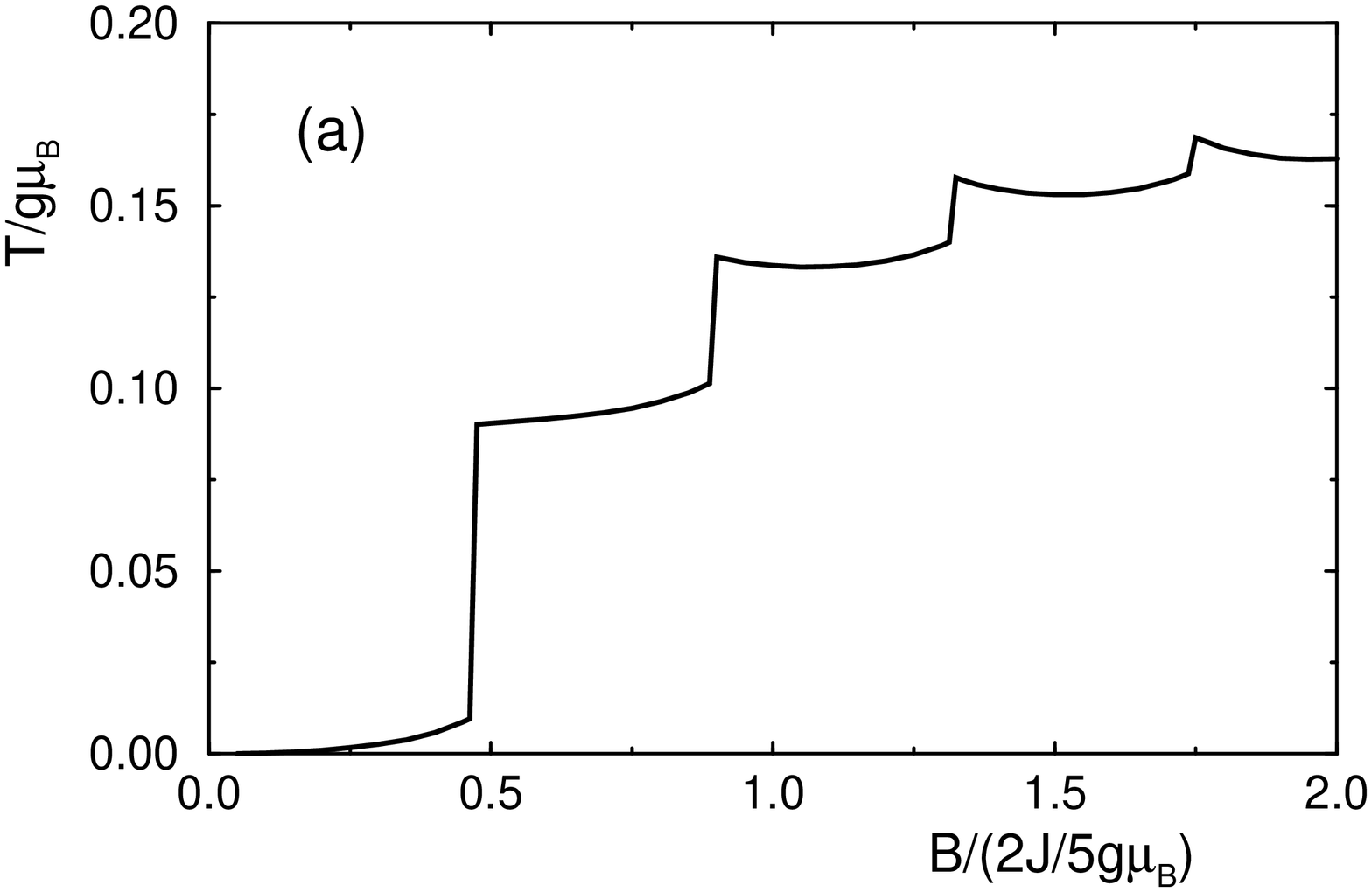,height=5cm,angle=270}}
\centerline{\psfig{figure=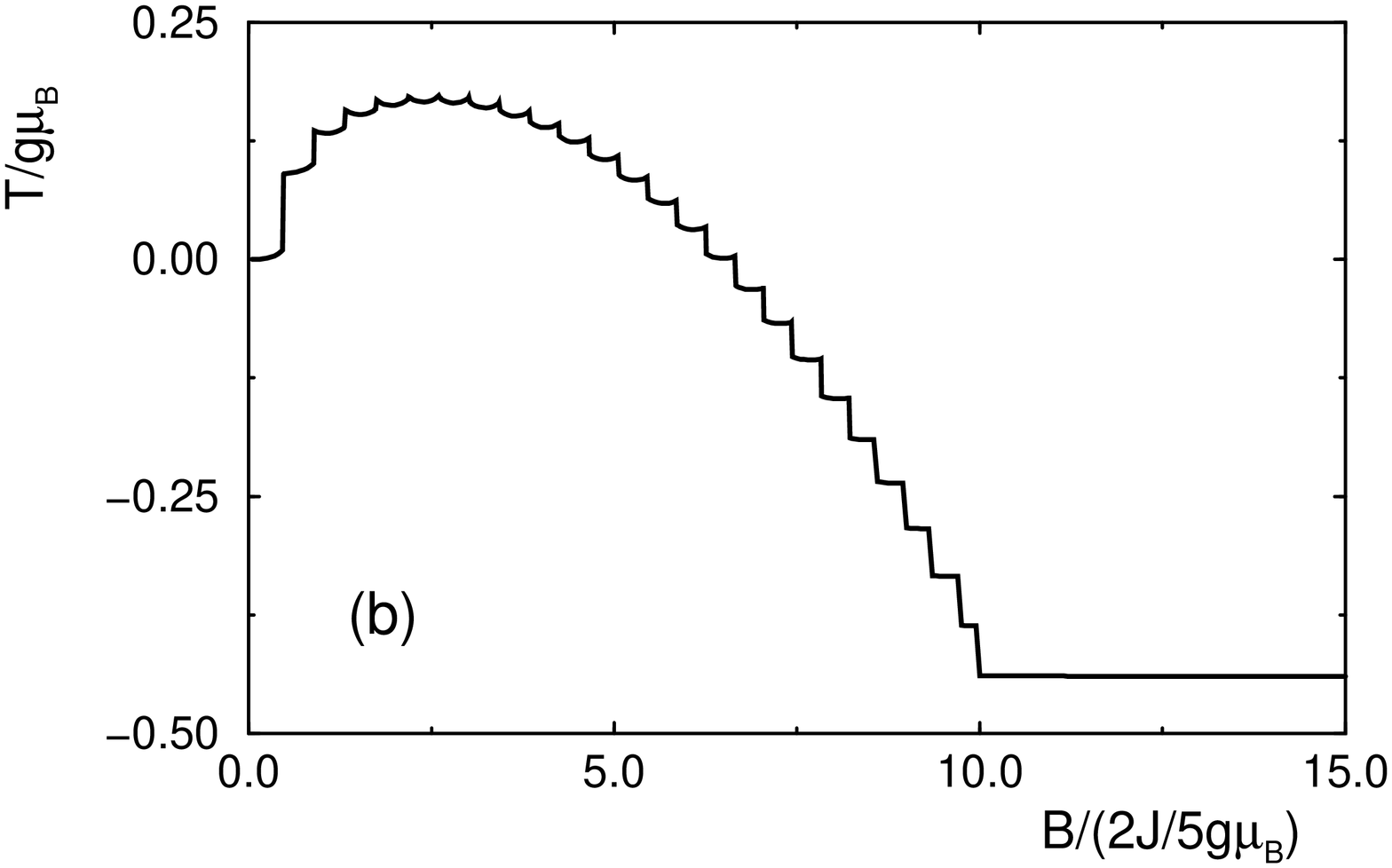,height=5cm,angle=270}}
\medskip
\caption{Torque curve of Fe$_{10}$ for $\psi = \pi/4$. (a) Detail at 
low fields; (b) full field range. } 
\end{figure}

With regard to the possible observation of quantum coherent tunneling 
of the N\'eel vector in ferric wheel systems, we may state that the 
magnetization and torque curves do appear to demonstrate that the 
desirable physical situation\cite{rcl} is realized. This consists of two, 
nearly-degenerate, low-lying energy levels, well separated from higher 
levels and weakly mixed by quantum (zero-point) transitions through a 
barrier, due to the easy-axis anisotropy, which separates them. However, 
the key property which cannot be shown by thermodynamic measurements 
such as the magnetization is that of coherence, or evolution of 
the quantum state over several oscillation cycles without scattering 
into a different state (decoherence). The experimental and theoretical 
aspects of this question, and the quantities (time-dependent correlation 
functions) which may be computed analytically and numerically to provide 
additional insight, will be addressed in a future publication.\cite{rmnl}

We conclude our analysis with a brief discussion of its relation to 
experiment. The existence of magnetization and torque steps, their 
locations for different field geometries, and their heights, are in good 
agreement with torque magnetometry results obtained\cite{rcja} for 
Fe$_6$ and Fe$_{10}$ samples at temperatures of 0.45K. We stress again 
that the simulations for the minimal model of Eq.~(\ref{emmh}) show a 
vertical step at the appropriate critical field for all field angles.
However, the recent, highly accurate results obtained\cite{raeau} for 
the first torque step in Na:Fe$_6$ at temperatures down to 30mK show two 
important features not reproduced by simulations of the model. One is 
the linear torque response observed for the $S = 0$ plateau, and the 
other a temperature-independent field broadening of the magnetization 
steps.

The first feature may be caused by admixtures of higher-spin states in 
the composition of the ground state, of either ``extrinsic'', meaning 
crystal misalignment or twinning, or ``intrinsic'' origin, meaning 
additional interaction terms not contained in (\ref{emmh}). Of yet 
more interest is the second feature. An intrinsic origin for this 
effect implies a term in the Hamiltonian permitting transitions of 
$\Delta S_z = \pm 1$. Such terms do not arise from biaxial anisotropy, 
particularly in a ring geometry. This in turn may indicate a possible 
role for superhyperfine interactions with proton spins on H sites 
neighboring the metallic ions. However, a study\cite{rcfpr} of this 
interaction in Fe$_{10}$ has found that its effects are too small by 
some orders of magnitude to explain the observed broadening. Similarly, 
hyperfine interactions with the Fe nuclear spins may be excluded because 
the natural abundance of $^{57}$Fe, the only species with non-zero nuclear 
spin, is so low. The most realistic intrinsic possibility is that the step 
width represents the energy scale of inter-molecular interactions, as any 
coupling between the moments of neighboring rings would broaden the 
magnetization response. The observed full width of the first step at 
40mK\cite{raeau} is 0.4T, which would correspond to an effective 
interaction of 0.14K per Fe atom in each ring. While the possible 
contributions to such interactions are difficult to quantify, simple 
geometrical considerations based on the crystallographic 
structure\cite{rccffggs} suggest that the long-range dipolar component 
could indeed be of this order. 

A candidate extrinsic origin\cite{raeau} is that molecules at the crystal 
surface may have different lattice parameters due to solvent loss, and 
thus undergo transitions at different field values corresponding to  
altered interaction strengths $J$. This effect should lead to rather 
small tails in the distribution of critical fields compared to the bulk 
response. Crystal mosaicity presents a further extrinsic possibility for 
step broadening, in that molecules with different angles to the applied 
field undergo transitions at different critical fields. In this scenario, 
the full width of the broadened first step at 40mK would correspond 
to a quasi-Gaussian angular distribution with a full width of 9.2$^0$, 
a rather large number which should be readily falsifiable from x-ray 
structural data. A more detailed analysis of these possibilities is 
deferred until further experimental data emerge to clarify the issue 
of extrinsic or intrinsic origin.

\section{Conclusion}

        We have presented a detailed analysis of the magnetization and 
torque response of the molecular magnetic systems Fe$_6$ and Fe$_{10}$. 
The step structure with critical fields $B_{cn}$, and variation of 
$B_{cn}$ with field angle to the ring plane, are readily understood from 
low-spin subspaces of a rigid-rotor Hamiltonian with weak spin anisotropy. 
Numerical simulations of the minimal but fully quantum mechanical model 
demonstrate clearly a vanishing step width, accompanied by rounding of 
the plateau regions even at zero temperature. These features can be 
related to those zero-point transitions permitted within this model 
when the system is close to magnetization steps, where energy 
levels of different spin states are in close proximity. The differences 
between simulation and experimental observation in Fe$_6$\cite{raeau} 
demonstrate the need for additional physical considerations on the 
theoretical side, and characterization of a wider range of samples on 
the experimental, with a view to discerning those features truly 
intrinsic to the ferric wheel systems. The angular variation of the step 
and rounding effects, which emerge from simulations and theory, should be 
observable in the most accurate experiments currently possible.\cite{raeau} 
These would be of particular interest in Fe$_{10}$, and may be expected 
to verify that the ferric wheel systems in transverse magnetic fields 
represent good candidates for the observation of quantum coherent 
tunneling of the N\'eel vector. 

\section{Acknowledgements}

We are indebted to M. Affronte, A. Cornia, and A. Jansen for invaluable 
discussion and provision of data. We are further grateful to D. Awschalom, 
F. Meier, D. Gatteschi, J. Harris, and R. Sessoli for helpful comments, 
and to the Swiss National Fund for financial support. BN acknowledges the 
generosity of the Treubelfonds, and the support of the Deutsche 
Forschungsgemeinschaft through SFB 484. XW and XZ acknowledge financial 
support from the University of Fribourg and the University of Neuch\^atel.

\end{document}